\documentclass[aps,pre,superscriptaddress,twocolumn]{revtex4-2}

\usepackage{amsmath}
\usepackage{amssymb}
\usepackage{graphicx}
\usepackage{color}
\usepackage{bm}
\usepackage{subfigure}
\usepackage{multirow}
\usepackage{latexsym}
\usepackage{mathptmx}
\usepackage{comment}

\begin{document}
\title{Patterning of multicomponent elastic shells by Gaussian curvature}

\author{Curt Waltmann}
\affiliation{Department of Materials Science and Engineering, Northwestern University, Evanston, IL, 60208}

\author{Ahis Shrestha}
\affiliation{Center for Computation and Theory of Soft Materials, Northwestern University, Evanston, IL 60208}
\affiliation{Department of Physics and Astronomy, Northwestern University, Evanston, IL, 60208}

\author{Monica Olvera de la Cruz}
\affiliation{Department of Materials Science and Engineering, Northwestern University, Evanston, IL, 60208}
\affiliation{Department of Chemistry, Northwestern University, Evanston, IL, 60208}
\affiliation{Department of Physics and Astronomy, Northwestern University, Evanston, IL, 60208}
\affiliation{Center for Computation and Theory of Soft Materials, Northwestern University, Evanston, IL 60208}

\begin{abstract}

Recent findings suggest that shell protein distribution and morphology of bacterial microcompartments regulate the chemical fluxes facilitating reactions which dictate their biological function. We explore how the morphology and component patterning are coupled through the competition of mean and Gaussian bending energies in multicomponent elastic shells that form three component irregular polyhedra. We observe two softer components with lower bending rigidities allocate on the edges and vertices while the harder component occupies the faces. When subjected to a non-zero interfacial line tension, the two softer components further separate and pattern into subdomains that are mediated by the Gaussian curvature. We find that this degree of fractionation is maximized when there is a weaker line tension and when the ratio of bending rigidities between the two softer domains $\approx 2$. Our results reveal a patterning mechanism in multicomponent shells that can capture the observed morphologies of bacterial microcompartments and, moreover, can be realized in synthetic vesicles.   
\end{abstract}

\maketitle

\section{Introduction} \label{sec:Intro}

The hierarchical nature of molecular biology allows for incredible complexity to arise from simple molecular building blocks. A striking example of this hierarchy is the division of biological processes into specific subcellular compartments \cite{Holthuis2013}. Although many of these subcomponents share similar building blocks such as lipids~\cite{vanMeer2008} and proteins~\cite{Schooch2004}, they have a vast range of properties~\cite{Funkhouser2013, Li2018, Schmit2021, Yuba2019, Hu2011, Vernizzi2007, Yuan2019, Brunk2020}, functions~\cite{Swift2013, Crowley2010, Yildiz2011, Oh2014}, and morphologies~\cite{Zandi2020, Greenfield2009, Leung2012, Sknepnek2011, Fanbo2023}. Thus, it is critical to understand how heterogeneity in the composition~\cite{Schmit2021, Girard2023, Jacobs2017} of these compartments can lead to unique morphologies and overall functions. We study a membrane compartment system composed of multiple components with different mechanical properties and investigate how these properties lead to complex surface patterning that affects shell morphology.

One class of subcellular compartments that are comprised of heterogeneous constituents is the bacterial microcompartment (BMC)~\cite{Kennedy2021, Kerfeld2018}. Similar to viruses, they have semi-permeable shells~\cite{Slininger2017} that are composed with copies of various shell proteins~\cite{Mills2021, Liu2021, Kennedy2021B, Kennedy2021J, Sutter2017, Lassila2014}. Atomistic simulations of these different shell protein have shown that there are unique mechanical interactions between the different pairs of shell proteins, namely, the bending interactions between two species of protein oligomers have distinct bending energies and energy-minimum bending angles~\cite{yaohua2021, Mills2021}. Thus, at a large-scale continuum level, the entire shell is expected to behave as a multicomponent membrane, particularly in terms of component-specific elastic parameters such as (mean) bending and Gaussian rigidity \cite{Baumgart2003,Julicher1993,Julicher1996,Sknepnek2012, SLi2021}. Understanding and quantifying the patterning mechanism in such multicomponent shells could provide key insights into how the different shell proteins are spatially arranged~\cite{Trettel2022} and their biological functions. Moreover, BMCs house enzymatic pathways that are involved in many other important cellular functions including $\text{CO}_{2}$ fixation. Hence, they are of great interest in the synthetic biology community due to their potential application as nano-bioreactors. For such applications, it is important to control the size and shape of these compartments as previous work~\cite{Mills2021} has shown that reaction rates can be altered by changing the morphology of BMCs. 

Fluid~\cite{Lipowsky1991, Seifert1997, Deserno2015} and crystalline~\cite{Seung1988, Kohyama2003, Bowick2001} membranes have been successfully studied using continuum elasticity theory that contain relatively few adjustable parameters~\cite{Helfrich1973, Winterhalter1988, LandauLifshitz}. For instance, viral capsids provide a great illustration of how buckling transitions in single component membranes are due to the competition of stretching and bending energies. Smaller virus particles tend to be spherical minimizing the membrane bending energy while larger particles buckle into icosahedra as the stretching energy scales with the surface area of the shell~\cite{Zandi2004, Lidmar2003}. In contrast, multicomponent membranes allow for a greater variety of shapes including irregular polyhedra~\cite{Vernizzi2011, Demers2012, Sknepnek2012, Yuan2019}. In particular, they form shapes with highly bent vertices and edges when one component is much less rigid than the others. Unlike icosahedral viruses studied in previous works~\cite{Zandi2004, Lidmar2003}, BMCs have been observed with irregular polyhedral shapes ~\cite{Kerfeld2018, Kennedy2020} possibly due to diverse mechanical properties that arise from heterogeneous bending interactions between the shell protein components~\cite{yaohua2021, Mills2021}. The question of how these components are patterned on the shell surface, which is highly relevant to functional features such as BMC permeability~\cite{Stewart2020, Slininger2017}, remains open~\cite{Trettel2022, Lassila2014}. In this work, we address the effects of component dependent rigidity as well as interfacial line tension on shell patterning and morphology.  

\begin{figure}[t!]
    \centering
    \includegraphics[width=1\columnwidth]{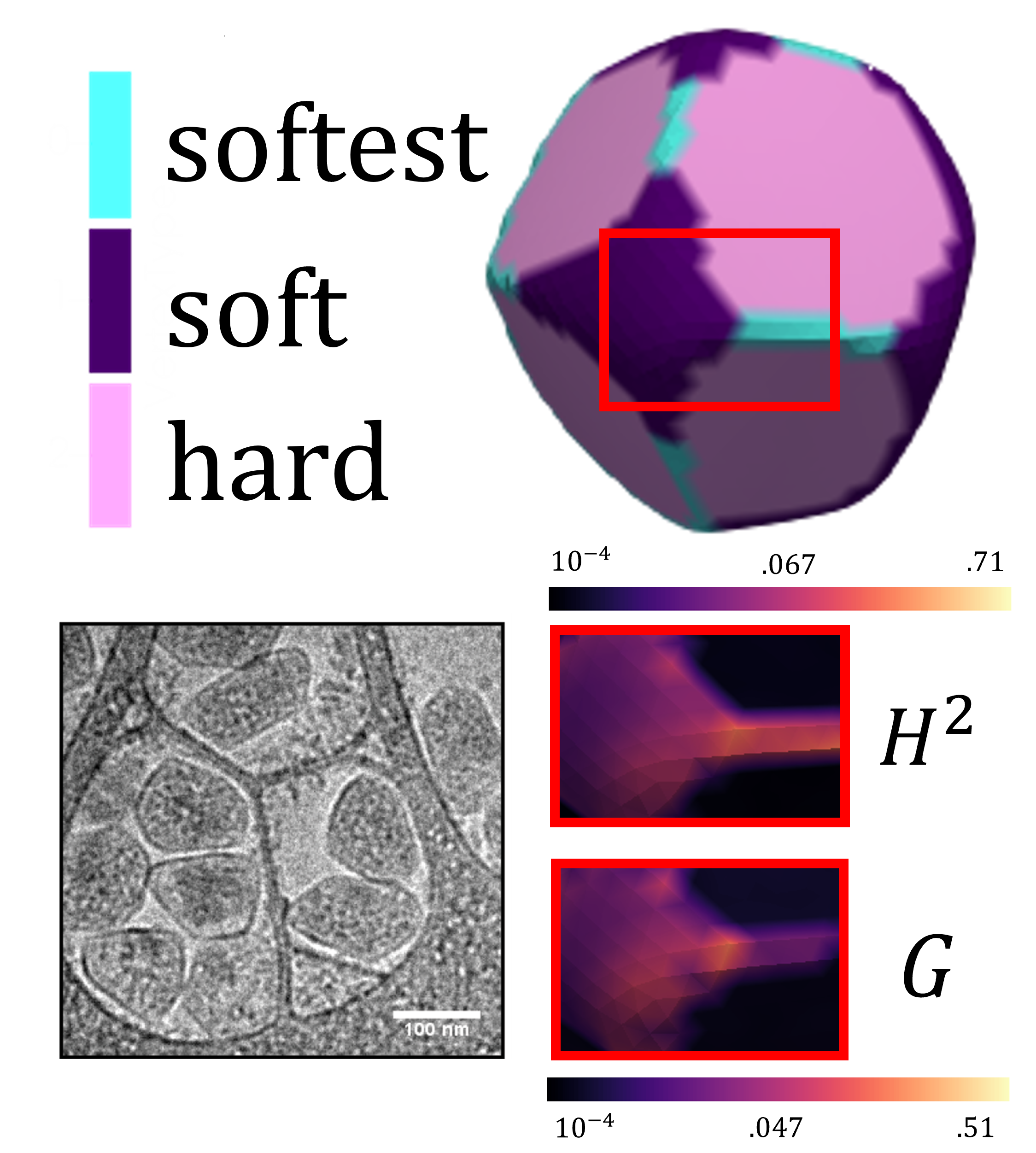}
    \caption{Multicomponent elastic shell with three distinct components denoted as hard (pink), soft (purple), and softest(cyan), referring to their different bending rigidities such that $\kappa_{hard}>\kappa_{soft}>\kappa_{softest}$ (top panel). These components pattern the shell based on the surface distribution of the square of twice the mean curvature $H^2$ and the Gaussian curvature $G$, which are given for the red box section of the shell with their magnitudes indicated by the color bar (lower right panel). This leads to polyhedra with inhomogeneous vertex curvature similar to cryo transmission electron micrographs of BMCs on TEM grids [Reproduced from Ref.~\cite{Kennedy2020} under  Creative Commons Attribution 4.0 International (CC BY) license] in the lower left panel.}
    \label{fig:intro}
\end{figure}

The role of Gaussian curvature becomes crucial for multicomponent membranes bearing domains of distinct rigidities ~\cite{Baumgart2003, Julicher1996, Vernizzi2011, Bassereau2018}. In a single component, uniform membrane, the Gaussian curvature is only relevant when considering a system that undergoes a change in topology. This can occur during processes such as membrane fusion or fission events, for instance, cell division or viral budding~\cite{RuizHerrero2015, Dharmavaram2019}. This is due to the Gauss-Bonnet theorem, by which the total Gaussian bending energy of a closed sphere amounts to a topological constant. However, in multicomponent membranes with components that have different bending and Gaussian rigidities, the total Gaussian energy of the system depends on the contact lines of the multidomain boundaries and the relative values of Gaussian rigidities in the domains \cite{Julicher1993, Julicher1996,Vernizzi2011, Bassereau2018, Baumgart2003, Shrestha2021}. The numerical value of the Gaussian rigidity of a given component has been shown to be directly proportional to its bending rigidity by the negative of a dimensionless constant whose reported value is approximately 1 in experiments and molecular dynamics simulation of lipid membranes~\cite{Hu2012}, as well as discretized mesh models~\cite{Seung1988, Vernizzi2007, Vernizzi2011, Sknepnek2012, Lidmar2003, Gompper1996, Schmidt2012}. As a result, a component with a larger bending rigidity will have a more negative Gaussian rigidity than the other components which can create a competition between the mean and Gaussian curvatures leading to patterning of components on the membrane. However, in previous works on multicomponent closed membranes, the mechanical regimes and conditions necessary to address these patterns have not been accessed~\cite{Sknepnek2012, Vernizzi2011, Hu2011, SLi2021, Demers2012}. Two component polyhedra~\cite{Sknepnek2012, Vernizzi2011} do not allow for multiple edge and vertex components needed for patterning, while three component icosahedra~\cite{SLi2021} inherently can not form the complex shapes dictated by this competition. Here, we investigate how the mean and Gaussian curvatures are distributed among two components which combine to form the edges and vertices of a buckled polyhedra whose faces are comprised of a third, much more rigid component (see Fig.~\ref{fig:intro}). This competition creates regions of high or low mean and Gaussian curvature which correspond to domains of the two softer components affecting the shell morphology. We investigate how this is impacted by the strength of a line tension between the different components, the ratio of bending rigidities between the components, and the overall composition of the membrane, finding values which maximize the number of domains. These low symmetry patterns give insight into how individual components may be distributed on subcellular compartments with complex and irregular shapes.

This article is arranged as follows. In Sec.~\ref{sec:Model}, we outline our continuum elasticity model and the numerical discretization methods we employ to minimize the total energy. Sec.~\ref{sec:Results} details our analysis of the resulting energy-minimum shell patterns under varied mechanical parameters. Our key findings and broader outcomes are highlighted in Sec.~\ref{sec:concl}. Additionally, we provide Appendices with supplemental materials and results.

\section{Model and methods}  \label{sec:Model}

\subsection{Continuum elasticity theory}
The elastic energy of thin shells is primarily written as the sum of mechanical stretching and bending contributions, in which the stretching term is given as
\begin{equation} \label{eq:ES}
    E_{Stretch} =  \frac{1}{2} \int_{S}  \frac{Y}{1 + \nu} \left(U(\Vec{r})_{ij}^{2} + \frac{\nu}{1+\nu}U(\Vec{r})_{kk}^{2} \right) dA  \,,
\end{equation}
where Y is the Young's modulus, $\nu$ is poisson's ratio, and $U(\Vec{r})_{ij}$ is the position $\Vec{r}$ dependent strain tensor~\cite{LandauLifshitz, Sknepnek2012, SLi2021}. The bending term is accounted by the Helfrich energy~\cite{Helfrich1973,Sknepnek2012, SLi2021} given by 
\begin{equation} \label{eq:EB}
    E_{Bend} = \frac{1}{2}\int_{S}  \kappa(\Vec{r}) (H(\Vec{r}) - H_{0}(\Vec{r}))^2 dA + \int_{S} \kappa_{G}(\Vec{r}) G(\Vec{r}) dA \,,
\end{equation}
where $\kappa(\Vec{r})$ is the bending rigidity, $H(\Vec{r})$ is twice the mean curvature, $H_{0}(\Vec{r})$ is the intrinsic or spontaneous curvature, $\kappa_{G}(\Vec{r})$ is the Gaussian rigidity, and $G(\Vec{r})$ is the Gaussian curvature. The mean and Gaussian curvature are functions of the two principal radii of curvature, $R_{1}(\Vec{r}) $ and $R_{2}(\Vec{r})$, such that $H = (\frac{1}{R_{1}} + \frac{1}{R_{2}})$ and $G = \frac{1}{R_{1}R_{2}}$. In the special case of perfect uniform sphere for instance, the position dependence in Eqs.~(\ref{eq:ES}) and~(\ref{eq:EB}) are lost. However, this dependence is maintained for more general multicompnent membranes with non-uniform shapes. The competition between the stretching and bending energies is characterized by the dimensionless Foppl von Karman number, Fvk$=\frac{YR^{2}}{\kappa}$ where $R$ is the shell radius. At high FvK values, the stretching term is dominant and a closed shell buckles into an icosahedron. Below the critical value FvK$\approx$ 154~\cite{Seung1988, Lidmar2003}, a smoother spherical shell remains with the bending energy density uniformly distributed. 

In this work, we investigate a system of three components (softest, soft, hard as shown in Fig.~\ref{fig:intro}) with different bending rigidities: $\kappa_{hard}$, $\kappa_{soft}$, and $\kappa_{softest}$. We define the ratios $\kappa_{soft}/\kappa_{softest} = \lambda_{1} \geq 1$ and $\kappa_{hard}/\kappa_{softest} = \lambda_{2}$. We set the ratio $\kappa_{hard}/\kappa_{soft} = 50$ such that $\lambda_2=50\lambda_1$, creating the right conditions for polyhedral buckling with the hard component on the faces~\cite{Vernizzi2011}. The final, softest component will have a bending rigidity $\kappa_{softest} \leq \kappa_{soft}$ and thus will buckle as well. Determining the exact value of the Gaussian rigidity of a given component, $\kappa_{G, \alpha}$, is not trivial and may dependent on the geometry as well as other intrinsic structural properties of the membrane system~\cite{Seung1988, Vernizzi2007, Vernizzi2011, Sknepnek2012, Lidmar2003, SLi2021, Deserno2015, Schmidt2012, Hu2012}. However, it is generally accepted that the Gaussian rigidity is proportional to the negative of the bending rigidity, such that $\kappa_{G, \alpha} = - c \kappa_{\alpha}$ where $c$ is a dimensionless positive constant. Moreover, we set zero intrinsic curvature $H_0=0$ in the membrane for simplicity. 

We now write the mean and Gaussian bending energy as a function of $\lambda_{1}$ and $\lambda_{2}$ by breaking up the shell into domains of the different components, $D_{\alpha}$. We reduce Eq.~(\ref{eq:EB}) as the dimensionless bending energy  
\begin{equation}
   E^*_{Bend} = E^*_{Mean}  -  c E^*_{Gauss} \,,
    \label{eq:c}
\end{equation}
where $E^*_{Bend}=E_{Bend}/\kappa_{softest}$, and the dimensionless mean and Gaussian energies are 
\begin{equation}
    E^*_{Mean} = \frac{1}{2} \int_{D_{softest}}{H^{2} dA} +  \frac{\lambda_{1}}{2} \int_{D_{soft}}{H^{2} dA} + \frac{\lambda_{2}}{2} \int_{D_{hard}}{H^{2} dA} \,,
    \label{eq:mean_lambda_1}
\end{equation}
and
\begin{equation}
    E^*_{Gauss}= \int_{D_{softest}}{G dA} + \lambda_{1} \int_{D_{soft}}{G dA} + \lambda_{2} \int_{D_{hard}}{G dA} \,,
    \label{eq:gauss_lambda_1}
\end{equation}
respectively. Eq.~\ref{eq:mean_lambda_1} implies that as $\lambda_{1}$ increases, the highest mean curvature regions should be occupied by the softest component to minimize the bending energy in Eq.~\ref{eq:c}. However, Eq.~\ref{eq:gauss_lambda_1} implies that the more rigid component should be located on those regions instead. Since, these regions also tend to be vertices of polyhedral shapes and thus there is a competition partitioning the mean and Gaussian curvature between the soft and softest components as a function of $\lambda_{1}$.

Given the presence of multiple, separated components at length scales where continuum modeling is reasonable, we impose an additional energy penalty when these components are in contact \cite{Lipowsky1992, Sknepnek2012, SLi2021, Dharmavaram2019, Shrestha2021}. To impose a penalty on the creation of an interface between two different components we add an interfacial line tension term \cite{Sknepnek2012, SLi2021} as follows 
\begin{equation}
    E_{Interface} = \Gamma_{\alpha, \beta} \int_{\partial S} dl \,,
\end{equation}
where $\Gamma_{\alpha, \beta}$ quantifies this energy penalty between two components, $\alpha$ and $\beta$. Thus, our total energy is $E = E_{Bend} + E_{Stretch} + E_{Interface}$.

\subsection{Discretized Monte Carlo simulations}

We discretize the shell surface into a triangular T=192 mesh in a spherical configuration with components placed on random mesh points according to the given fractions of each component~\cite{Sknepnek2012, SLi2021, Vernizzi2011}. It has a radius of approximately 11.5$r_{0}$ where $r_{0}$ is the equilibrium length between vertices. We then minimize the discretized version of the total elastic energy as developed by Seung and Nelson~\cite{Seung1988} with the components on the mesh points in order to include the line tension term~\cite{Sknepnek2012} using a Monte Carlo protocol. We attempt two different kinds of Monte Carlo moves. One type moves a mesh point 5\% of the equilibrium mesh spacing and the second swaps the identity of two mesh points. This allows us to minimize the morphology and patterning of the membrane. 

In the Seung-Nelson (SN) discretization, we assign harmonic springs between each pair of nearest neighbor vertices~\cite{Sknepnek2012, SLi2021, Vernizzi2011}. When the components are defined on the vertices, this gives us the SN discretized stretching energy as  
  \begin{equation}
     E^{SN}_{Stretch} = \frac{1}{2} \sum_{i, j} \epsilon_{\alpha, \beta} (|\Vec{r}_{i} - \Vec{r}_{j}| - r_{0})^2   \,,
 \end{equation}
where $\alpha$ and $\beta$ are the identities of vertices i and j respectively and when $\alpha \neq \beta$ one must define a mixing rule for the harmonic spring constant $\epsilon_{\alpha, \beta}$. However, in practice, we set $\epsilon_{\alpha, \beta}=\epsilon$, as a constant. We also set $r_0=1$, the unit of length. In the continuum limit this potential leads to a Young's modulus, $Y=\frac{2}{\sqrt{3}}\epsilon$ and Poisson ratio, $\nu = \frac{1}{3}$ ~\cite{Seung1988}.

The SN discretized bending energy is computed in terms of the angle between triangles which share edge i,j where $k_{\alpha}$ is the discrete bending rigidity, and $n_{i}$ and $n_{j}$ are the unit normal vectors of triangles $t_{i}$ and $t_{j}$, for which there is dihedral angle, $\theta_{i,j}$. Here, the components are defined on the vertices instead ~\cite{Sknepnek2012, SLi2021, Vernizzi2011} and the expression becomes 
\begin{equation}
    E^{SN}_{Bend} = \frac{1}{2} \sum_{t_{i},t_{j}} k_{\alpha, \beta} |\hat{n_{i}} - \hat{n_{j}}|^{2} = \sum_{t_{i},t_{j}} k_{\alpha, \beta} (1 - \cos(\theta_{i,j})) \,,
\end{equation} 
where $k_{\alpha, \beta}$ is the discrete bending rigidity of an edge connecting vertices with identities $\alpha$ and $\beta$. We choose the mixing rule to be an average of the two discrete bending rigidities $k_{\alpha, \beta} = \frac{k_{\alpha} + k_{\beta}}{2}$. We note that such a parameter is not defined the continuum modeling which only defines the bending rigidity as a property of only one component. Here and in the single component system, the relation between the discrete bending constant, k, and the continuum bending rigidity, $\kappa$, is assumed to be $\kappa = \sigma k$ with $\sigma = \frac{\sqrt{3}}{2}$ following the works~\cite{Seung1988,Lidmar2003,Schmidt2012}. For this discretization, we estimate that $c \approx 1.28$ by using it as a fitting parameter for our data (see Appendix~\ref{appen:Est}). This is in good agreement with previous works~\cite{Lidmar2003,Schmidt2012} which show that c=$\frac{4}{3}$ is a valid solution for this mesh for both spheres and cylinders. We choose the discrete bending rigidities such that FvK$_{hard}\approx 18$, FvK$_{soft}\approx 900$, and  FvK$_{softest} \approx  900 \lambda_{1}$. 

Following previous work of Sknepnek et al.~\cite{Sknepnek2012}, we define the discrete interfacial line tension as a sum over the neighboring mesh points i,j
\begin{equation}
    E^{Sknepnek}_{Interface} = \frac{1}{2}\sum_{i,j} \gamma_{\alpha,\beta} (1 - \delta_{\alpha,\beta}) \,,
\end{equation}
where $\delta_{\alpha,\beta}$ is the Kronecker delta and $\gamma_{\alpha, \beta}$ is the discrete line tension parameter with units of energy. Thus, it only penalizes the number of interfacial contacts and not the interface length as in the continuum model and it can not lead to domain budding. The factor of $\frac{1}{2}$ is included to avoid double counting these contacts. Here, we make $\gamma_{\alpha, \beta}=\gamma$ a constant such that all interfaces are penalized equally. Throughout this work, we show the magnitude of $\gamma$ relative to the bending energy using $\frac{\gamma}{k_{hard}}$ where $k_{hard}$is the discrete bending constant for the hard component.

\section{Results and discussion} \label{sec:Results}

\begin{figure}[h]
    \centering
    \includegraphics[width=1\columnwidth]{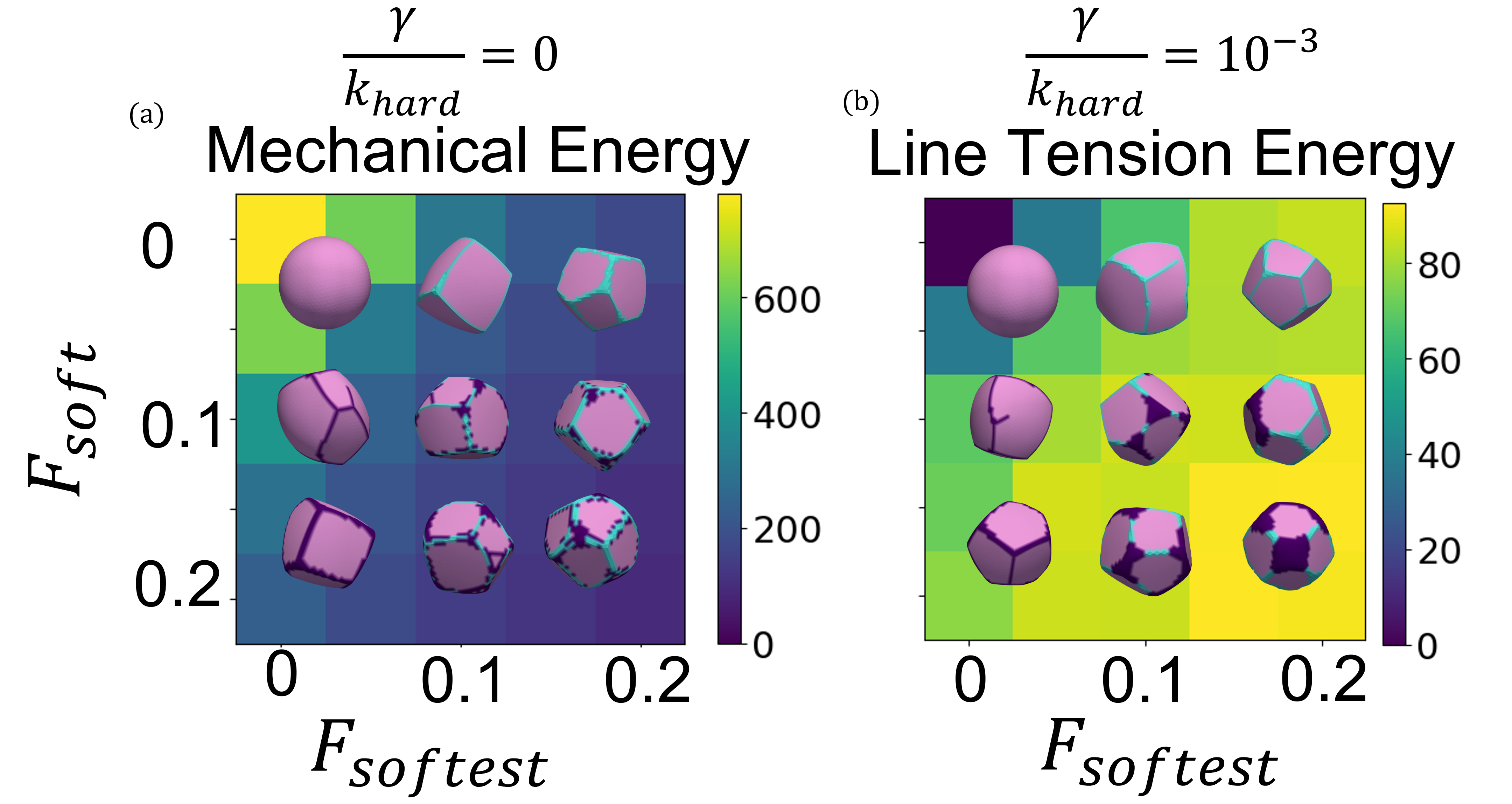}
    \caption{ Typical energies and shell configurations with and without line tension, where the rigidity ratio, $\frac{\kappa_{soft}}{\kappa_{softest}} = \lambda_{1} = 2$.  (a) The mechanical energy, $E_{Bend} + E_{Stretch}$, for elastic shells with different fractions of the hard (pink), soft (purple), and softest (cyan) components. $F_{Hard}$, the fraction of the hard component is given by $1 - F_{softest} - F_{soft}$. Characteristic shell patterning is shown for all combinations where $F_{soft} = 0.0, 0.1, 0.2$ and $F_{softest} = 0.0, 0.1, 0.2$. (b) The line tension energy, $E_{Interface}$ for an elastic shell with line tension. Characteristic shell patterning is shown for all combinations where $F_{soft} = 0.0, 0.1, 0.2$ and $F_{softest} = 0.0, 0.1, 0.2$.    }
    \label{fig:energies_both}
\end{figure}

Membrane morphologies are shown for different component fractions of the softest and soft components, $F_{softest}$ and $F_{soft}$ respectively, with and without line tension in Fig.~\ref{fig:energies_both} along with the corresponding energies. These configurations are generated after an annealing protocol, which we explore in the Appendix. We note the mechanical energy, that is the sum of only the stretching and bending energies, drops quickly as softer components are added to a one component shell made only of the hard component. This accompanies a polyhedral buckling transition as shown in previous work~\cite{Sknepnek2012, Vernizzi2011}. However, once a certain amount of the softer components is added ($F_{softest} + F_{soft} > 0.1$) the mechanical energy seems to slowly approach a minimum. This seems to signify that the buckling transition is "complete" (Fig.~\ref{fig:energies_both}(a)). A related trend is seen in the line tension energies, which increase as more of the softer components are added in Fig.~\ref{fig:energies_both}(b). The line tension energy linearly for smaller $F_{soft} + F_{softest}$ as adding the softer components creates a line or edge that has an associated energy with it. However, at the point where this buckling transition is completed, the relationship between the amount of the softer components and the line tension energy is no longer linear and instead the line tension increases much more slowly. At this point, the bending energy is no longer dominating and the line tension energy can be optimized as well.

\begin{figure}[htbp]
    \centering
    \includegraphics[width=1\columnwidth]{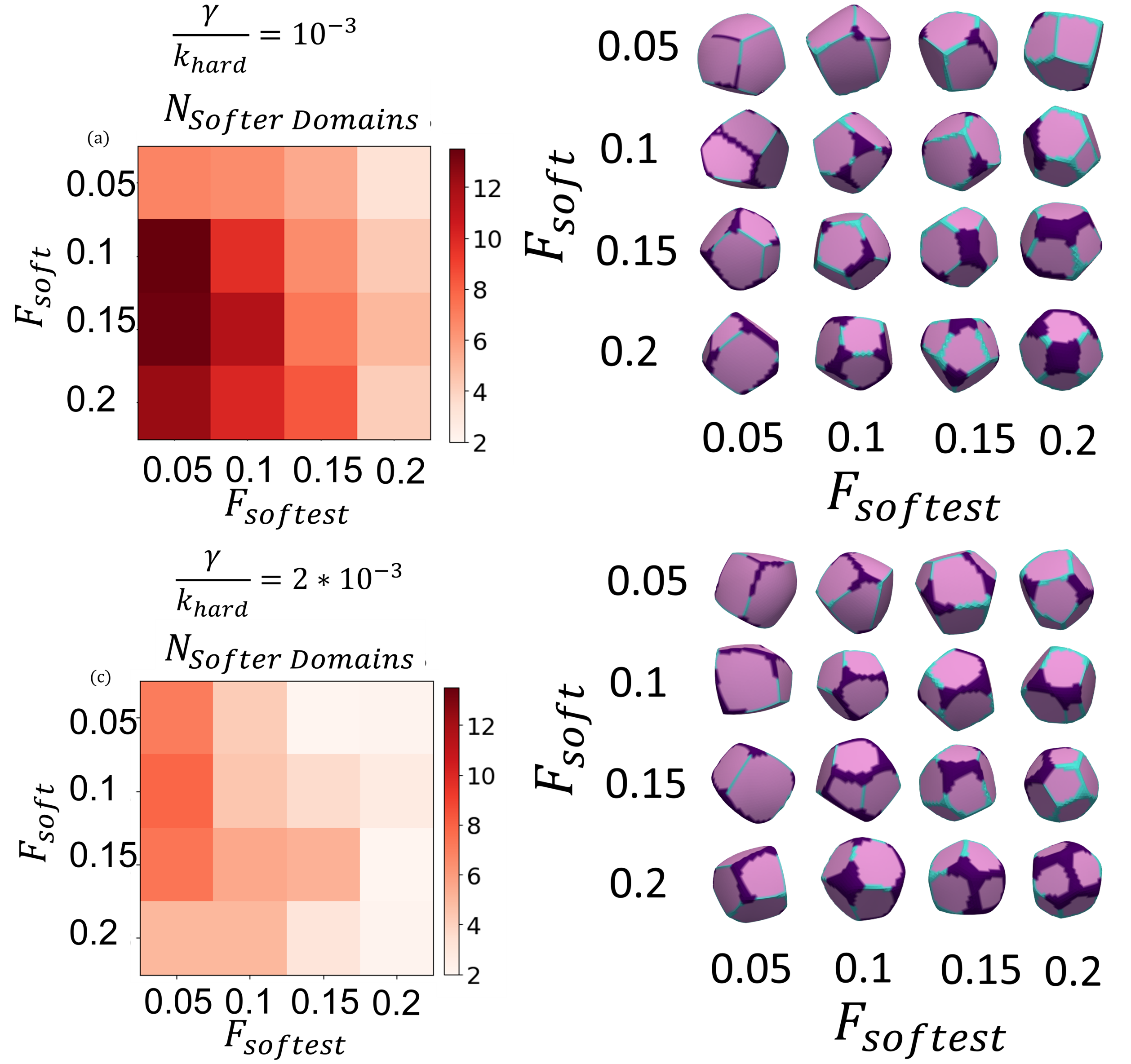}
    \caption{Analysis of domain patterning for different compositions when the rigidity ratio, $\lambda_{1}=2$.  (a) The number of softer domains for different fractions the soft (purple) and softest (cyan) components for the line tension parameter, $\frac{\gamma}{k_{hard}}$ = $10^{-3}$. A softer domain is a set of mesh points of either the soft or softest component where, for all mesh points in the set, there exists a path between mesh points that is only connected by neighboring mesh points of that same component. Representative snapshots for the same component fractions and line tension parameter are shown in (b). (c) The number of softer domains for different fractions the soft (purple) and softest (cyan) components for the line tension parameter, $\frac{\gamma}{k_{hard}}$ = 2* $10^{-3}$.  Representative snapshots for the same line tension parameter are shown in (d).}
    \label{fig:domains_pd}
\end{figure}

\begin{figure}[t!]
    \centering
    \includegraphics[width=1\columnwidth]{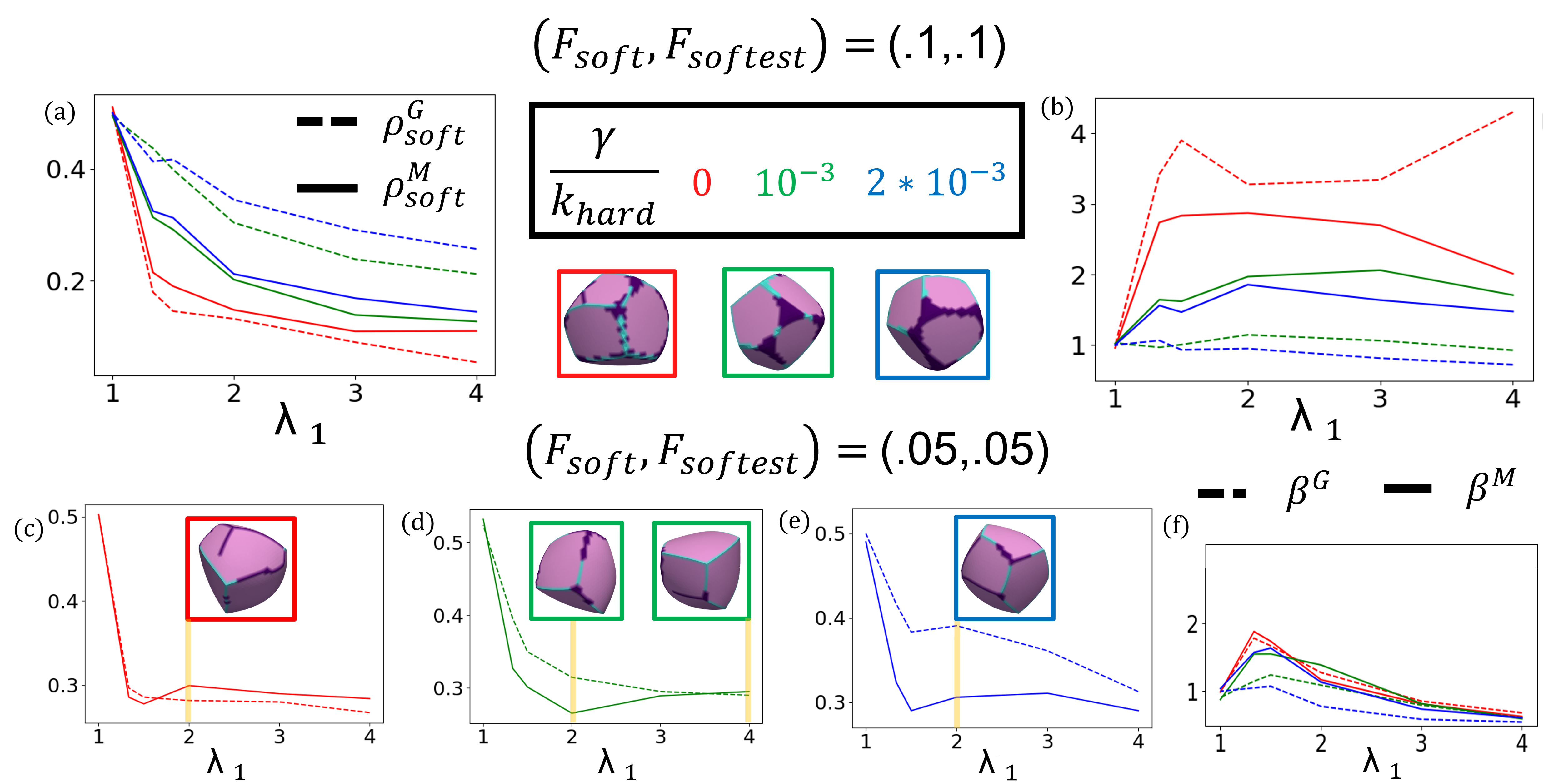}
    \caption{Analysis of the partitioning of the mean and Gaussian curvatures also as a function of the rigidity ratio, $\lambda_{1} = \frac{\kappa_{soft}}{\kappa_{softest}}$. The different colors are for different line tension parameters, $\frac{\gamma}{k_{hard}}$. Dashed lines are the fraction of the Gaussian curvature found on the soft component (purple), $\rho^{G}_{soft}$, as defined in eq.~\ref{eq:row_g_soft}. Solid lines are the fraction of the mean curvature found on the soft component, $\rho^{M}_{soft}$, as defined in eq.~\ref{eq:row_m_soft}. (a) Relative to the mean curvature, the Gaussian curvature is partitioned preferentially to the soft component ($\rho^{G}_{soft} >\rho^{M}_{soft}$) for $F_{soft}=F_{softest}=0.1$. This only occurs in the presence of a line tension, but doubling the line tension parameter has little effect on this partitioning but does lead to domain fusion. For $F_{soft} = F_{Softest}=0.05$, (c) when no line tension, there is only a slight preference for the Gaussian curvature to be on soft domains ($\rho^{G}_{soft} > \rho^{M}_{soft}$) at low values of $\lambda_1$, with mean squared curvature dominating for other $\lambda_1$ ($\rho^{G}_{soft} < \rho^{M}_{soft}$). (d) When weak line tension is added there is a preference for the Gaussian curvature to partition to the soft component, but only when $\lambda_{1}>3$. (e) When the line tension is doubled there is always a preference for the Gaussian curvature to be located on the soft component, although the preference is maximized at $\lambda_{1}=2$. The ratio of mean and Gaussian curvature energies between the softest and soft components, $\beta^G$ and $\beta^M$, as defined in eq.~\ref{eq:beta_g} and \ref{eq:beta_m} respectively for different values of line tension when (b) $F_{soft}=F_{softest}=0.1$ and (f) $F_{soft}=F_{softest}=0.05$.}
    \label{fig:lambda_lt}
\end{figure}

When the line tension is present, we observe segregated domains of the softer components in Fig.~\ref{fig:energies_both}(b), while without the line tension, we observe patterns that look like the soft component (purple) is "reinforcing" the softest component (cyan) which is located at the center of the edges. In Fig.~\ref{fig:domains_pd}, we quantify the number of these softer domains $N_{Softer Domains}=N_{Soft Domains}+N_{Softest Domains}$, that is, the sum total of the domains which contains both the soft and the softest parts. More precisely, a softer domain is defined as a set of mesh points of either the soft or softest component where, for all vertices in the set, there exists a path between mesh points that is only connected by neighboring mesh points of that same component. Fig.~\ref{fig:domains_pd}(a) shows that the number of softer domains is maximized when the fraction of the soft component, $F_{soft}$ is higher then that of the of softest component, $F_{softest}$, but when the total amount of softer components is not too high ($F_{soft} + F_{softest} \leq 0.25$). Representative morphologies shown in Fig.~\ref{fig:domains_pd}(b) suggest that the soft (purple) component prefers to be located on the vertices of the polyhedra, while the softest (cyan) component tends to be located on the edges. This creates the right conditions for many domains to form since the soft vertices will be separated from each other by the softest  edges. Adding more of the soft component fuses the vertex domains leading to a decrease in the number of independent domains. The soft vertex domains also seem thicker than the cyan edges, leading to inhomogeneous mean curvature of the vertices and edges based on which component the edge or vertex is made of. Another way to fuse the domains should be to increase the magnitude of the line tension parameter, $\frac{\gamma}{k_{hard}}$. This penalizes the formation of independent domains. Indeed, Fig.~\ref{fig:domains_pd}(c) shows a decrease in the number of domains for almost every composition by doubling $\frac{\gamma}{k_{hard}}$. The one composition which seems unaffected by the increased line tension is $F_{soft} = F_{softest} = 0.05$, which is addressed later in the text and in Fig.~\ref{fig:lambda_lt}.

\begin{figure}[h]
    \centering
    \includegraphics[width=1\columnwidth]{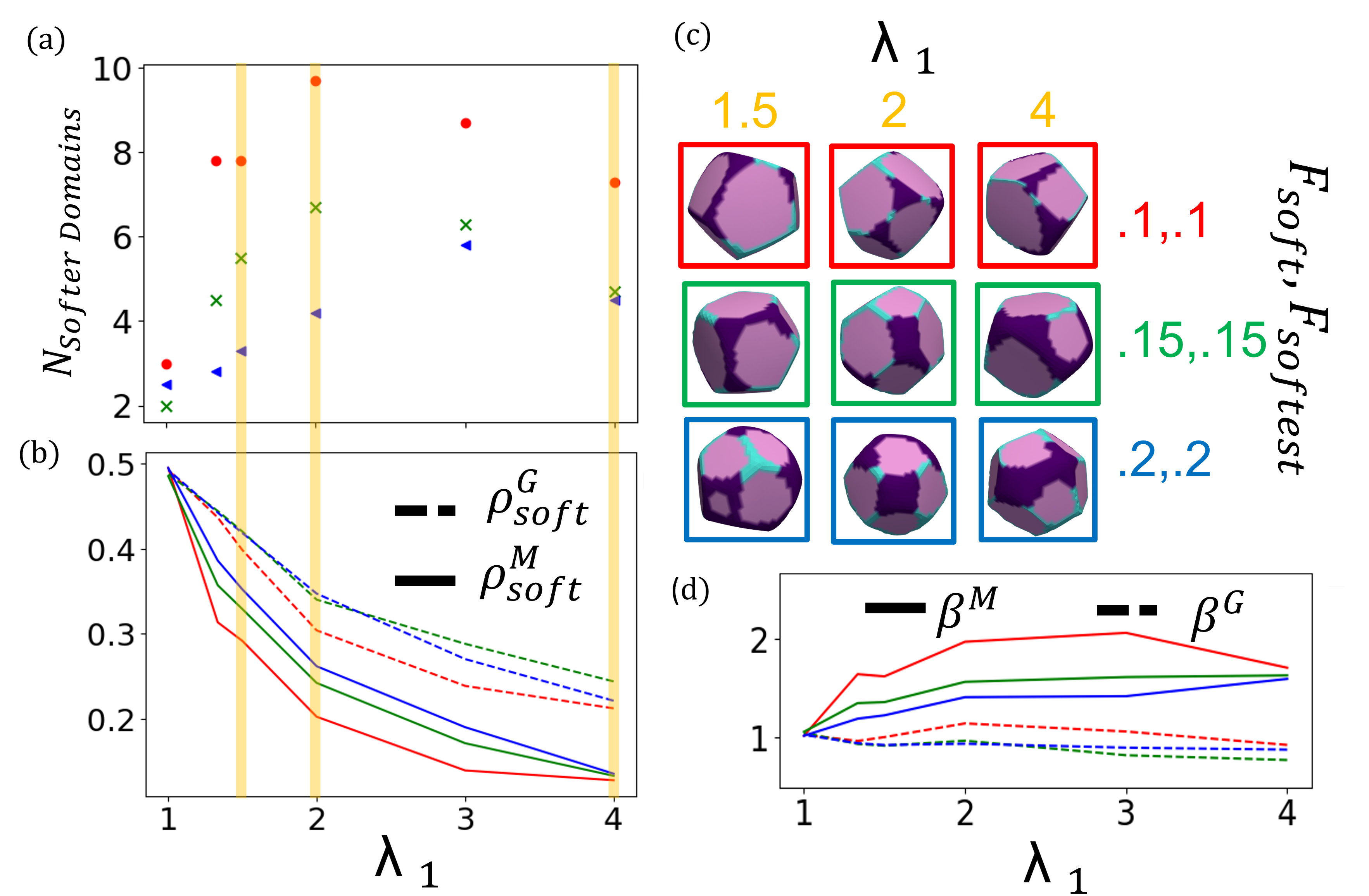}
    \caption{Analysis of the relationship between domain formation and the distribution of the mean and Gaussian curvatures. All values shown are with line tension paramter $\frac{\gamma}{k_{hard}} = 10^{-3}$ (a) The number of softer domains as a function of the rigidity ratio, $\lambda_{1} = \frac{\kappa_{soft}}{\kappa_{softest}}$ for three different compositions shown in (c). (b) Analysis of the partitioning of the mean and Gaussian curvatures also as a function of the rigidity ratio, $\lambda_{1}$. Dashed lines are the fraction of the Gaussian curvature found on the soft component, $\rho^{G}_{soft}$, as defined in eq.~\ref{eq:row_g_soft}. Solid lines are the fraction of the mean squared curvature found on the soft component, $\rho^{M}_{soft}$, as defined in eq.~\ref{eq:row_m_soft}. There is more Gaussian and mean squared curvature on the softest component (cyan), however, there is a higher fraction of Gaussian curvature on the soft component than mean curvature on the soft component. (c) Snapshots of different compositions and different $\lambda_{1}$, which correspond to (a) and (b). (d) The ratio of Gaussian and mean curvature energy on the softest and softest components, $\beta^G$ and $\beta^M$, as defined in eq~\ref{eq:beta_g} and \ref{eq:beta_m}, respectively. }
    \label{fig:lambda_domains}
\end{figure}

Snapshots in Fig.~\ref{fig:domains_pd}(d) visually show the fusion of soft and softest domains. Although the preference for the soft component to be located on the vertices is still noticeable, there are now more soft edges and these edges are not as clearly defined as their softest (cyan) counterparts. We expect that the persistent preference of the soft component to be located on the vertices is due to the Gaussian curvature of the membrane, which is concentrated at the vertices of the polyhedra. The relation between the Gaussian and bending rigidity imples that the soft component has a more negative Gaussian rigidity despite having a more positive bending rigidity. Thus, the soft component should be preferred in regions of high Gaussian curvature such as the vertices of the polyhedra, as we observe. Based on our definition of the Gaussian energy in eq.~\ref{eq:gauss_lambda_1}, this relationship should also be a function of $\lambda_{1}$, the ratio between the rigidity of the soft and softest components. To illustrate this, we vary the rigidity ratio in the range $1 \leq \lambda_1 \leq 4$, tracking the number of domains for different compositions. We also quantity the Gaussian and mean contributions located on the soft component with respect to the softest component through the following dimensionless quantities  
\begin{equation}
    \rho^{G}_{soft} = \frac{\int_{D_{soft}} G dA }{ \int_{D_{Soft}} G dA + \int_{D_{softest}} G dA}
    \label{eq:row_g_soft}
\end{equation}
and 
\begin{equation}
    \rho^{M}_{soft} =  \frac{\int_{D_{soft}} H^2 dA }{ \int_{D_{Soft}} H^2 dA + \int_{D_{softest}} H^2 dA}
    \label{eq:row_m_soft}
\end{equation}
These is computed using the per node definitions of the Gaussian and mean curvatures found in eq.~\ref{eq:vertex_mean} and \ref{eq:vertex_gauss}. In Fig.~\ref{fig:lambda_domains}(a), we observe a maximum in the number of softer domains for rigidity ratios of 2 or 3, depending on the specific compositions. In Fig.~\ref{fig:lambda_domains}(b), we show that both $\rho^{G}_{soft}$ and $\rho^{M}_{soft}$ are always less than .5, meaning that more of both the mean and Gaussian contributions are on the softest component. However, there is always a higher Gaussian contribution than mean contribution on the soft component ($\rho^{G}_{soft} >\rho^{M}_{soft}$). This creates a maximum in the number of domains when the rigidity ratio is large enough for there to be a difference in Gaussian rigidity of the two components but not so large that the mean curvature term dominates completely. This is illustrated by the snapshots at different $\lambda_{1}$ values in Fig.~\ref{fig:lambda_domains}(c). As $\lambda_{1}$ increases the softest component creates edges that are skinnier relative to the soft component allowing the softest component to cover more of the surface, increasing its fraction of the Gaussian and mean curvatures as shown in Fig.~\ref{fig:lambda_domains}(b). At first, this leads to more segregated domains as it mostly occupies the edges, leaving segregated vertices occupied by the soft component. However, it eventually takes up more of the vertices as well, leading to a decrease in the average number of domains.

One can approximate the relationship between $\rho^{G}_{soft}$ and $\lambda_1$ goes as $1 / (1 + \beta^G \lambda_1)$, where $\beta^G$ can be a fitting parameter that describes how fast $\rho^{G}_{soft}$ decays. It can be shown through Eq.~(\ref{eq:beta_g}) that in this approxtimation, we find
\begin{equation}
    \beta^G = \frac{ \int_{D_{softest}} G dA}{ \lambda_{1} \int_{D_{Soft}} G dA} \,,
    \label{eq:beta_g}
\end{equation}
which means that here $\beta_G$ is the ratio of Gaussian energy in the softest component to that of the soft component. Similary, for the mean curvature using Eq.~(\ref{eq:beta_m}), we get
\begin{equation}
    \beta^M = \frac{ \int_{D_{softest}} H^2 dA}{ \lambda_{1} \int_{D_{Soft}} H^2 dA} \,,
    \label{eq:beta_m}
\end{equation}
Interestingly, $\beta^G \approx 1$ for all values of $\lambda_{1}$ and for different compositions as shown in Fig.~\ref{fig:lambda_domains}(d), which means the soft and softest components have the same Gaussian curvature energy. This is not the case for the mean curvature energy, where $\beta^{M}$ has a stronger dependence on $\lambda_1$ and seems to plateau around 1.6 for all compositions. This means that the softest component is taking on more of the mean curvature energy. The difference in $\beta^G$ and $\beta^M$ quantifies the energies involved in the morphological "splitting" of the mean and Gaussian curvatures through the low mean curvature vertices of the soft component.

Finally, we look at the effect of line tension on the splitting of the mean and guassian curvatures. In Fig.~\ref{fig:lambda_lt}(a), we plot $\rho^{G}_{soft}$ and $\rho^{M}_{soft}$  as a function of the rigidity ratio at $\frac{\gamma}{k_{hard}} = 0, 10^{-3}, 2* 10^{-3}$ and $F_{soft} = F_{softest} = 0.1$. We observe only a slight difference in the curves for  $\frac{\gamma}{k_{hard}} = 10^{-3}$  and $2 * 10^{-3}$ with a slight increase in the fraction of both mean squared and Gaussian curvature on the soft component at higher values of $\lambda_{1}$. This suggests that the magnitude of the line tension plays only small role in determining how the mean and Gaussian curvatures are distributed between the two components, despite the fact that the lower value of line tension lead to significantly more domains in Fig.~\ref{fig:domains_pd}. The values of $\beta^G$ and $\beta^M$ also remain nearly unchanged and can be seen in Figure~\ref{fig:lambda_lt}(b). Thus, this effect was only due to domain fusion and not a change in curvature distribution or energy. For shells without line tension, patterning on the vertices is lost as there is more mean squared curvature than Gaussian curvature on the soft component and both $\beta^G$ and $\beta^M$ become much larger with $\beta^G > \beta^M$ (see Fig.~\ref{fig:lambda_lt}(f)). This is the case when there is plenty of excess softer component to buckle the shell as shown in Fig. ~\ref{fig:energies_both}. In contrast, when all of the softer components are needed to buckle the shell, as in Fig.~\ref{fig:lambda_lt}(b-d) where $F_{soft}=F_{softest}=0.05$, the line tension has a large impact on the partitioning of mean and Gaussian curvature. Even without line tension in Fig.~\ref{fig:lambda_lt}(c), there is a slight preference for the Gaussian curvature on the soft phase when $1< \lambda_1 < 2$. When the weak line tension is applied, this preference is much stronger and extends to $\lambda_{1}>3$ (Fig.~\ref{fig:lambda_lt}(d)). When the line tension is doubled there is always a preference for the Gaussian curvature to be located on the soft component, although the preference is maximized at $\lambda_{1}=2$ and decreases afterwards as shown in Fig.~\ref{fig:lambda_lt}(e). Here the line tension and rigidity ratio seem to be competing against each other, with line tension favoring the formation of soft domains located on the vertices. 

\section{Conclusion and Outlook} \label{sec:concl}

In this article, we show that Gaussian curvature plays a crucial role in the morphology and patterning of multicomponent elastic membranes. We observe, for shells composed of three components with distinct bending rigidities, energy-minimum configurations where the two softer components distributed on the edges and vertices while the harder component forms the faces. Among the two softer components that occupy the edges and vertices, a preference of the component with higher bending rigidity is predominantly found at the vertices leading to many subdomains, although with low symmetry. These vertices were much flatter than their counterparts occupied by the component with lower bending rigidity leading to polyhedra with inhomogeneous vertices as has been seen in BMCs with many components. These effects required that the two softer components on the edges and vertices have different bending rigidities but not so different that one component will occupy the vast majority of the bent edges and vertices. 

We saw a peak in the number of domains at a rigidity ratio of $\approx 2$, in contrast with the larger ratios necessary for the polyhedral buckling transition. Since this ratio is relatively low, the necessary differences in bending properties may result from only subtle changes to chemical properties of the molecules that make the different shell proteins. This is consistent with the BMC system where the two most common types of shell proteins found are structurally similar and have almost identical amino acid sequences~\cite{Kennedy2021J}. These effects also required a line tension to help segregate the components, but not one so strong as to fuse the domains back together, again suggesting relatively similar components can create these morphological patterns. 

Besides such distinctions in their elastic properties, the shell protein components of BMCs are also known to have distinct electrostatic properties, namely, they carry different ratios of charged amino acids. Thus, the resulting morphology and component patterning observed here may in turn lead to patterned surface charge densities on the shell as seen in the icosahedral shells that have been crystallized. Moreover, such BMC systems can facilitate chemical reaction that produce fluxes of ionic chemical species. The combined effect of such surface features provided with their asymmetric distribution on the shell can induce self-phoretic movement~\cite{Kamat2022, Shrestha2023}. Both the elastic and electrostatic properties arise due to small compositional variations in the protein components, and thus, it may also be possible to regulate these subtle chemical perturbations in synthetic vesicles that can be optimized for specialized functionalities.

\begin{acknowledgements}
CW was funded by the Northwestern University Nicholson Fellowship. AS was supported by the Department of Energy (DOE), Office of Basic Energy Sciences under Contract DE-FG02-08ER46539. MOdlC thanks the computational support of the Sherman Fairchild Foundation. CW, AS and MOdlC thank the support of the Center for Computation and Theory of Soft Materials. 
\end{acknowledgements}

\appendix

\section{Itzykson Discretization} \label{appen:Itzyk}
In the main text, we use the SN dicretization~\cite{Seung1988} of the Helfrich bending and stretching energy, but define the components on the vertices instead of the edges~\cite{Sknepnek2012}. This makes it simple to define the line tension, however, the edges can now have a mixed identity at the interface and one must define mixing rules, which are somewhat arbitrary and do not exist in the continuum equations. One could also use the Itzykson discretization~\cite{Kohyama2003, SLi2021}. Here, the components are located on the vertices, however the bending energies are computed by calculating the mean and Gaussian curvatures at the vertex instead of the angle across triangles. In Fig.~\ref{fig:si_itz_hot}, we show that this discretization is does not perform well for our annealing protocol and thus we do not show results using the Itzykson discretization in the main text. However, we do use the definitions of mean and Gaussian curvature at a vertex to analyze our morphologies generated by the SN discretization.

The issue of mixed edges is resolved in the Itzykson~\cite{Kohyama2003} discretization by computing the local mean and Gaussian curvature of the Voronoi cell around each vertex. The potential, summed over all vertices, takes on the form
\begin{equation}
    E_{Bend} = \sum_{i} \left(\frac{\kappa_{\alpha}}{2} (H_{i} - H_{0, \alpha})^2 + \kappa_{G, \alpha}G_{i} \right) A_{i} \,,
\end{equation}
where $\kappa_{\alpha}$ and $\kappa_{G, \alpha}$ have the same value as the continuum parameters and thus we use the same $\kappa$ to denote them as opposed to the SN discretization, which uses k that must be converted to $\kappa$. $H_{i}$ and $G_{i}$ are the mean and Gaussian curvature at vertex i and $A_{i}$ is the dual-lattice vertex area defined as 
\begin{equation}
    A_{i} = \frac{1}{8} \sum_{j \in N(i)} [\cot(\phi_{i,j}) + \cot(\psi_{i,j})] (r_{i} - r_{j})^{2} \,,
\end{equation}
where the sum is over the neighbors of i, indexed by j. The angles $\phi_{i,j}$ and $\psi_{i, j}$ are the interior angles opposite from edge i,j which is shared by triangles, $t_{i}$ and $t_{j}$. The mean curvature is computed as 
\begin{equation}
H_{i} = \frac{sign}{2A_{i}} \sum_{j \in N(i)} [\cot(\phi_{i,j}) + \cot(\psi_{i,j})](r_{i} - r_{j}) \,,
\label{eq:vertex_mean}
\end{equation}
where $sign$ denotes whether the curvature is concave or convex, 1 or -1 respectively. The Gaussian curvature is found by summing over the angles adjacent to vertex i in the triangles which share vertex i as follows 
\begin{equation}
    G_{i} = \frac{1}{A_{i}} \left( 2\pi - \sum_{t} \theta_{t} \right)  \,,
\label{eq:vertex_gauss}
\end{equation}
where $\theta$ is the angle at vertex i. In Figure~\ref{fig:si_itz_hot}, we show this discretized form of the energy is not adequately minimized on our mesh by our Monte Carlo protocol. However, we use this discretization to calculate the mean and Gaussian curvature of the mesh minimized with the SN discretization in Figures~\ref{fig:lambda_domains} and \ref{fig:lambda_lt} in the main text.

\section{Estimation of $c$ and $\sigma$} \label{appen:Est}

\begin{figure}[h]
    \centering
    \includegraphics[width=1\columnwidth]{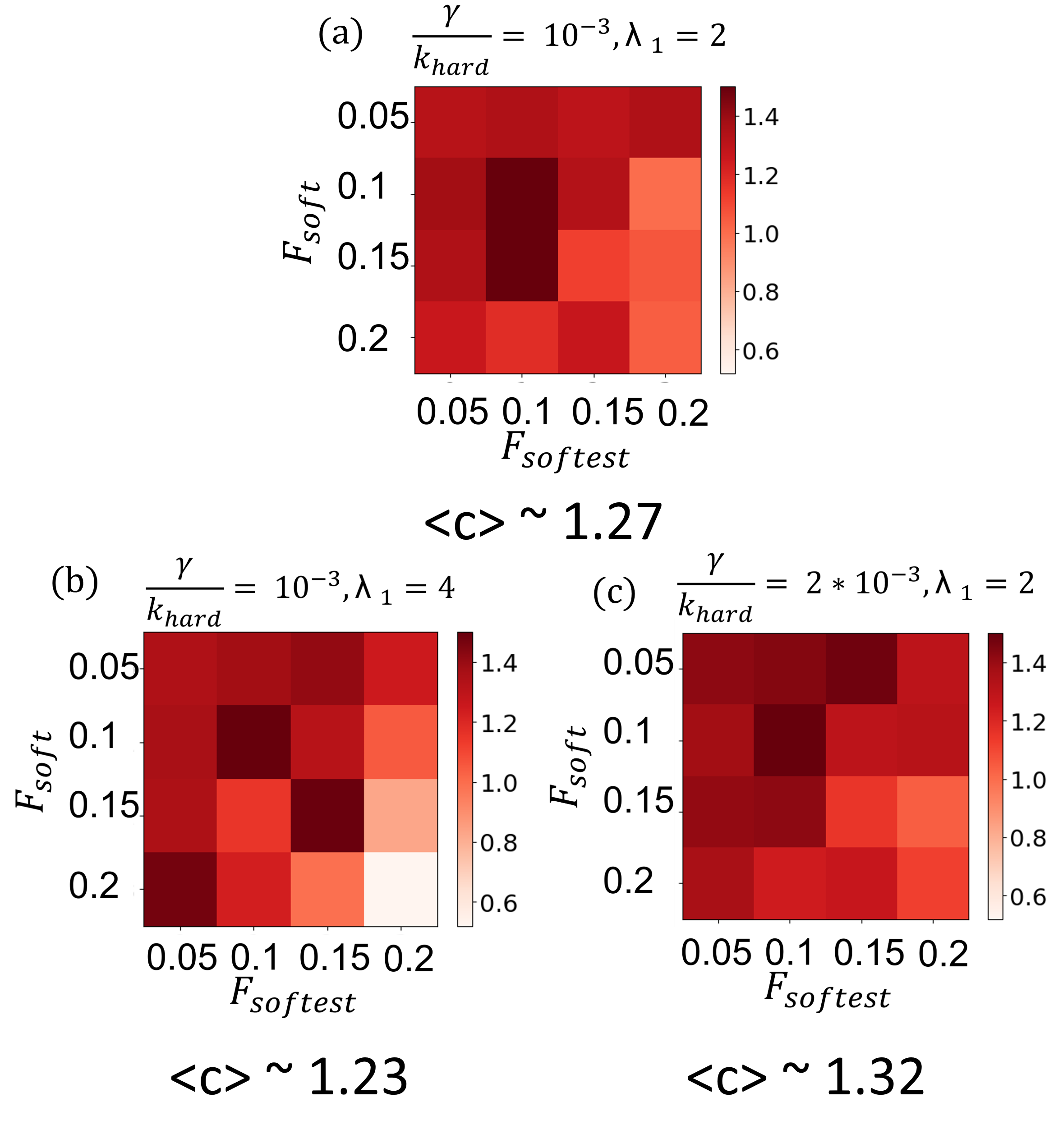}
    \caption{Using a fitting scheme where the continuum solution is $c=1.28$ as described in the main text we estimate c for different compositions and the values of $\frac{\gamma}{k_{hard}}$ and $\lambda_{1}$ shown in (a), (b), (c). We also display the average value of c for the different compositions at the bottom of (a), (b), and (c) which also show reasonable agreement with the continuum solution.}
    \label{fig:si_c}
\end{figure}

We attempted to estimate the continuum $c$ and $\sigma$ for our three component meshes in a way that is consistent with the continuum result for the one component sphere. In the SN discretization of a one component sphere, the discrete bending energy is $4\pi k \frac{\sqrt{3}}{3}$. We can set this equal to the corresponding continuum bending energy
\begin{equation}
    4\pi k \frac{\sqrt{3}}{3} = \frac{\kappa}{2} \int H^2 dA - c \kappa \int G dA = 4 \pi \kappa (2-c) \,,
\end{equation}
and thus 
\begin{equation}
     \kappa=\frac{\sqrt{3}}{3(2-c)}k = \sigma k  \,.
\end{equation}

We will use this relationship to reduce c and $\kappa$ to a single fitting parameter. Lidmar et al.~\cite{Lidmar2003} previously used the $\kappa = \frac{\sqrt{3}k}{2}$ for a cylinder with no Gaussian curvature to arrive at the result that $c=\frac{4}{3}$; a solution which is correct for both geometries. We use 
\begin{equation}
    E^{discrete}_{Bend} = \kappa_{softest} E^*_{Bend} \,,
\end{equation}
where $E^{discrete}_{bend}$ is the discretized bending energy of the mesh and $E^*_{Benc}$ is the same as defined in eq.~\ref{eq:c}. We found that the best fit for the data, including different compositions, line tensions, and rigidity ratios was $c \approx 1.28$ and $\sigma \approx 0.80$. This is in fairly good agreement with the values suggested by Lidmar et al. In Figure~\ref{fig:si_c}, we show c for some specific parameters.

\section{Monte Carlo annealing} \label{appen:Anneal}

\begin{figure}[!htbp]
    \centering
    \includegraphics[width=1\columnwidth]{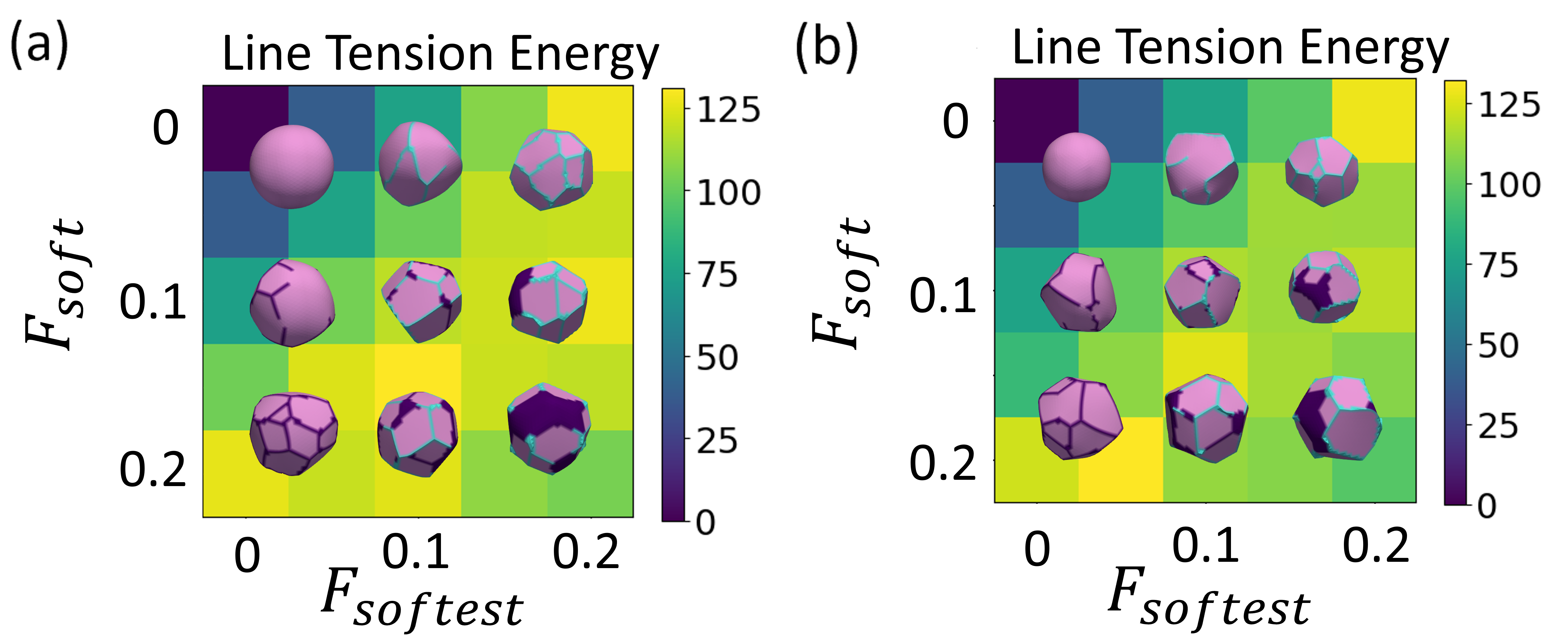}
    \caption{ Results for annealing using (a) Seung-Nelson discretization and (b) Itzykson discretization, both with a lower maximum temperature, $T^*=0.01$. Typical energies and shell configurations with and without line tension, where the rigidity ratio, $\frac{\kappa_{soft}}{\kappa_{softest}} = \lambda_{1} = 2$ and $\frac{\gamma}{k_{hard}} = 10^{-3}$. The line tension energies and characteristic shell patterns are shown for all combinations where $F_{soft} = 0.0, 0.1, 0.2$ and $F_{softest} = 0.0, 0.1, 0.2$.}
    \label{fig:line_tension_comb}
\end{figure}

\begin{figure}[!htbp]
    \centering
    \includegraphics[width=0.75\columnwidth]{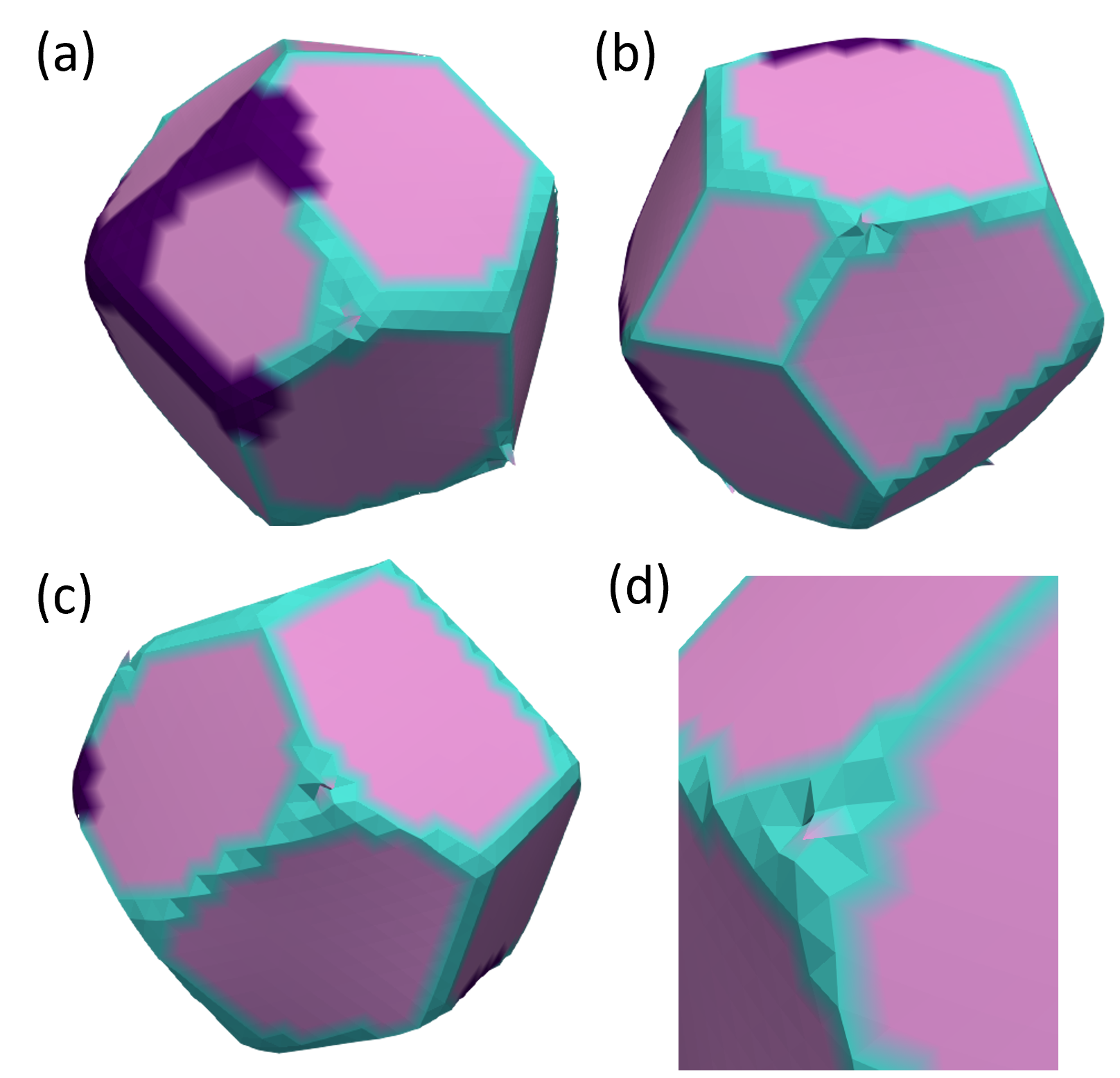}
    \caption{One representative shell morphology for annealing using the Itzykson discretization with the standard maximum temperature, $T^*=0.1$. The rigidity ratio, $\frac{\kappa_{soft}}{\kappa_{softest}} = \lambda_{1} = 2$, $\frac{\gamma}{k_{hard}} = 10^{-3}$, and $(F_{softest}, F_{soft}) = (0.2, 0.1)$. Instabilities at vertices composed of the softest (cyan) component in (a)-(c) with (a) showing some preference of the soft component to occupy the vertices. (d) Shows a magnified image where it is apparent that some of the hard component is trapped and there are many convex bending angles. }
    \label{fig:si_itz_hot}
\end{figure}

Our annealing protocol involves 7 different temperature cycles. The first 5 cycles start at an annealing temperature, $T^{*}$, of $0.1$ and then is reduced linearly in 10 steps to $T^{*}=10^{-7}$. The 6th and 7th cycles start at $T^{*}=0.01$ and $T^{*}=0.001$ respectively before being cooled to $T^{*}=10^{-7}$ in the same linear fashion over 10 steps. For each $T^*$, we run 5000 Monte Carlo steps for a total of 350000 Monte Carlo steps. In each Monte Carlo step, we attempt two different kinds of Monte Carlo moves. (1) For each vertex, we attempt to move it a distance of $0.05r_{0}$ and (2) For each vertex, we attempt to swap the identity of two random vertices. The probability of move acceptance, $P$, is given by 
\begin{equation}
    P = min \left( 1, e^{-\Delta E /T^{*}} \right) \,,
\end{equation}
where $\Delta E$ is the difference in total energy of the system after the attempted Monte Carlo move. This protocol is run 4 separate times with different random initial conditions for all sets of parameters reported in the main text. All values reported are an average of these 4 trials. Using a maximum $T^{*}=0.1$ is important as if a similar protocol is used but the maximum $T^*=0.01$, the shells end up in much higher energy states as shown in Fig.~\ref{fig:line_tension_comb}(a). This is for the same parameters as in Fig.~\ref{fig:energies_both}(b) in the main text and one can see the much higher line tension energies in these cases as well as a lack of smooth trends in Fig.~\ref{fig:line_tension_comb}(a).  

We also ran this annealing protocol using the Itzykson discretization and achieved very similar results in Fig.~\ref{fig:line_tension_comb}(b). Neither Fig.~\ref{fig:line_tension_comb}(b) nor Fig.~\ref{fig:line_tension_comb}(a) show lowest energy states do the lower maximum annealing temperature, $T^{*}$. When we attempted to use the higher annealing temperature, $T^{*}=1$, with the Itzykson discretization we found that the mesh became unstable as shown in Fig.~\ref{fig:si_itz_hot}. This issue appears at vertices formed by the softest component, making it impossible to quantitatively compare the Itzykson and Nelson discretizations for our parameters. We do note that the soft component still seems to prefer the vertices as was the case with the Nelson discretization in the main text. This is noticeable in Fig.~\ref{fig:si_itz_hot}(a).   

\section{Larger line tension} \label{appen:Line}

\begin{figure}[b!]
    \centering
    \includegraphics[width=0.6\columnwidth]{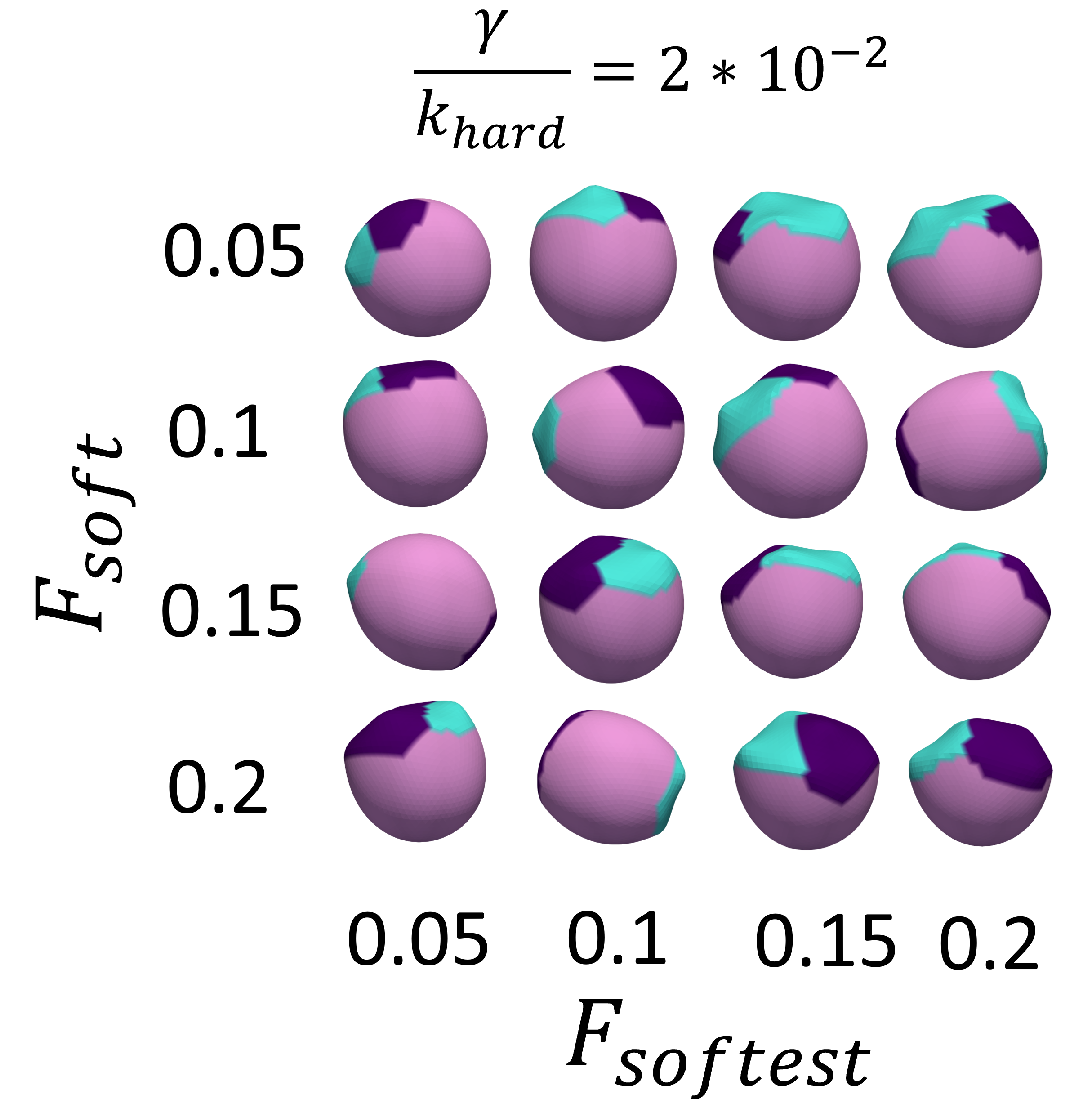}
    \caption{ Representative snapshots for the shells with a strong line tension parameter. The rigidity ratio, $\lambda_{1}=2$. There are always 2 softer domains which are now completely separated from the hard component and the polyhedral shape is lost. Similar effects have been seen previously in a two component system with high line tension~\cite{Sknepnek2012}. }
    \label{fig:si_lt10}
\end{figure}
In the main text, we show that the line tension is necessary for patterning the polyhedron, but also that doubling the line tension will fuse domains back together. In Fig.\ref{fig:si_lt10}, we show that as the line tension is increased by an order of magnitude, the polyhedral morphology is lost altogether and a sphere with segregated regions of the softer component are formed instead. A very similar transition has been observed in two component membranes~\cite{Sknepnek2012}. This also supports the idea that line tension in these systems must be relatively weak or else the polyhedral morphology will not be observed.

\bibliographystyle{apsrev4-2}
\bibliography{citations}

\end{document}